\documentclass{article}

\usepackage[margin=1in]{geometry}
\usepackage{amsmath,amssymb,amsthm}
\usepackage[normalem]{ulem} 
\usepackage{mathtools}
\usepackage{bm}
\usepackage{enumitem}
\usepackage{graphicx}
\usepackage{hyperref}
\usepackage{algorithm}
\usepackage{algpseudocode}
\usepackage{tikz}
\usetikzlibrary{arrows.meta,shapes.geometric}

\makeatletter
\renewcommand\@fnsymbol[1]{}
\makeatother

\title{Game-Theoretic Drone Swarm Defense\\[0.5em]\large A Case Study in Applied Differential Game Theory}
\author{Ross E. Allen\\[0.4em]\normalsize MIT Lincoln Laboratory%
    \thanks{\\DISTRIBUTION STATEMENT A. Approved for public release. Distribution is unlimited.\\
        This material is based upon work supported by the Under Secretary of War for Research and Engineering under Air Force Contract No. FA8702-15-D-0001 or FA8702-25-D-B002. Any opinions, findings, conclusions or recommendations expressed in this material are those of the author(s) and do not necessarily reflect the views of the Under Secretary of War for Research and Engineering.\\
        \textcopyright{} 2026 Massachusetts Institute of Technology. \\
        Delivered to the U.S. Government with Unlimited Rights, as defined in DFARS Part 252.227-7013 or 7014 (Feb 2014). Notwithstanding any copyright notice, U.S. Government rights in this work are defined by DFARS 252.227-7013 or DFARS 252.227-7014 as detailed above. Use of this work other than as specifically authorized by the U.S. Government may violate any copyrights that exist in this work.}
}
\date{August 2026}

\begin{document}

\maketitle

\section*{Executive Summary}

This technical report is a study of the use of differential game (DG) theory to solve the target-assignment and midcourse guidance problems of drone swarms tasked with intercepting opposing swarms in defense of high-value assets. %
The game-theoretic tactics---which treat the intruder swarm as a rational agent and seek a Nash equilibrium between defenders and intruders---are compared against baseline tactics that model the defense problem as a unilateral optimization of the defenders' maneuvers. %
Monte Carlo simulation and Bayesian analysis show that the game-theoretic approach has a higher probability of successfully intercepting all intruders than the baseline techniques. %
This improvement in successful defense probability is most pronounced when the intruder swarm is capable of evasive maneuvers: relative to baseline optimization tactics, differential-game tactics increase estimated defense success from 94.6\% to 96.8\%, closing approximately 41\% of the remaining gap to perfect defense. %
To add statistical credibility to this result, a paired-trial Bayesian analysis assigns a 99.9\% posterior probability that differential-game tactics have a higher probability of successful asset defense than baseline tactics in this scenario.

\clearpage
\tableofcontents
\clearpage

\section{Problem Formulation}
\label{sec:problem_formulation}

The drone swarm defense problem is posed as a many-agent variant of the Lady-Bandit-Guard problem described by Rusnak~\cite{rusnak2005lady}. %
A team of $N_G$ guard drones (also referred to as \emph{defenders} or \emph{interceptors}) is tasked with intercepting an incoming swarm of $N_B$ bandit drones (also referred to as \emph{intruders} or \emph{threats}) before any of them reach the set of $N_L$ high-value assets (HVAs, called ``ladies'' in Rusnak's terminology~\cite{rusnak2005lady}). 

\begin{figure}[t]
\centering
    \begin{tikzpicture}[
        >=Stealth,
        scale=1.05,
        guard/.style={circle, fill=blue!65, draw=blue!80!black,
                    minimum size=7pt, inner sep=0pt},
        bandit/.style={circle, fill=red!75, draw=red!80!black,
                    minimum size=7pt, inner sep=0pt},
        ghost bandit/.style={bandit, opacity=0.35},
        lady/.style={star, star points=5, fill=green!60!black,
                    draw=green!40!black, minimum size=12pt, inner sep=1pt},
        guard path/.style={blue!70!black, line width=1.1pt, ->},
        bandit path/.style={red!75!black, line width=1.1pt, ->},
        faint guard path/.style={blue!70!black, line width=0.8pt,
                                opacity=0.23, ->},
        faint bandit path/.style={red!75!black, line width=0.8pt,
                                opacity=0.23, ->},
        candidate path/.style={blue!70!black, line width=0.8pt,
                            dashed, opacity=0.35, ->}
    ]

    \coordinate (G) at (-3.1,-1.5);
    \coordinate (B) at (-0.5,4.1);
    \coordinate (L) at (0.6,0.1);
    \coordinate (I) at (-0.25,2.0);
    \coordinate (B1) at (-2.55,3.65);
    \coordinate (B2) at (0.95,3.50);
    \coordinate (I1) at (-1.35,1.55);
    \coordinate (I2) at (1.05,1.55);

    \draw[bandit path]
        (B)
        .. controls (-0.15,3.35) and (-0.85,2.55) ..
        (I)
        .. controls (0.45,1.25) and (0.90,0.70) ..
        (L);

    \draw[guard path]
        (G)
        .. controls (-1.25,-0.45) and (-1.30,1.60) ..
        (-0.90,2.15)
        .. controls (-0.70,2.60) and (-0.35,2.55) ..
        (-0.42,2.30)
        .. controls (-0.48,2.15) and (-0.55,2.25) ..
        (I);

    \draw[faint bandit path]
        (B1) .. controls (-2.15,2.80) and (-1.95,2.05) .. (I1);
    \draw[faint bandit path]
        (B2) .. controls (1.25,2.75) and (1.25,2.05) .. (I2);

    \draw[candidate path]
        (G) .. controls (-3.05,0.10) and (-2.45,1.35) .. (I1);
    \draw[candidate path]
        (G) .. controls (-1.25,-0.55) and (0.35,0.55) .. (I2);

    \node[guard]  at (G) {};
    \node[bandit] at (B) {};
    \node[lady]   at (L) {};

    \node[ghost bandit] at (B1) {};
    \node[ghost bandit] at (B2) {};
    \node[ghost bandit] at (I1) {};
    \node[ghost bandit] at (I2) {};

    \node[ghost bandit] at (I) {};

    \node[below left=2pt]  at (G) {$G_i$};
    \node[above=2pt]       at (B) {$B_j$};
    \node[below right=2pt] at (L) {$L_{\ell}$};


    \end{tikzpicture}
\caption{Conceptual drone swarm defense problem posed as a many-agent Lady-Bandit-Guard problem~\cite{rusnak2005lady}. The guard drone $G_i$, which is one of a large set of guard drones, attempts to intercept a particular bandit drone $B_j$ before it reaches a high-value asset $L_{\ell}$, while considering the trade-offs of potentially pursuing different bandits.} %
\label{fig:swarm_defense_concept}
\end{figure}
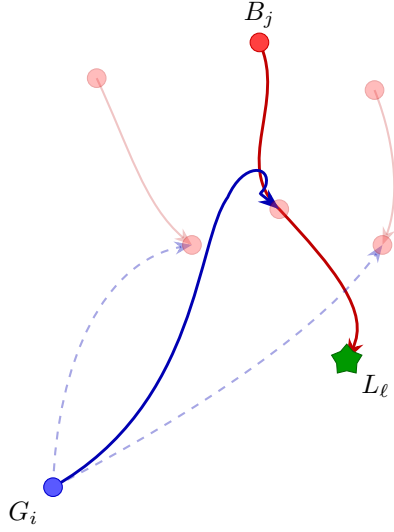

The problem is multi-faceted, generally requiring optimization of both the assignment of specific guards to intercept specific bandits and guidance laws that generate dynamically-feasible trajectories that achieve those interceptions. Fig.~\ref{fig:swarm_defense_concept} gives a conceptual illustration of an engagement with a small number of drones.

\subsection{Teams, State, and Information}
\label{sec:teams_state_information}
\label{sec:vehicle_dynamics}

Let $G\triangleq\{0,\ldots,N_G-1\}$, $B\triangleq\{0,\ldots,N_B-1\}$, and $L\triangleq\{0,\ldots,N_L-1\}$ index the guards, bandits, and high-value assets (HVAs), respectively. The total population sizes $N_G$, $N_B$, and $N_L$ are fixed for an episode. At any time, $G_A\subseteq G$ and $B_A\subseteq B$ denote the active guard and bandit index sets, with $N_{G_A}\triangleq|G_A|$ and $N_{B_A}\triangleq|B_A|$. Inactive vehicles remain represented in the full environment state but are omitted from active-roster planning problems.

Guards are modeled as fixed-speed, planar Dubins vehicles. For guard $i\in G$, the state, control, and equations of motion are
\begin{equation}
\begin{aligned}
\mathbf{p}_{G,i}&=\begin{bmatrix}p_{x,G,i} & p_{y,G,i}\end{bmatrix}^{\mathsf T},
& \mathbf{x}_{G,i}&=
\begin{bmatrix}\mathbf{p}_{G,i}^{\mathsf T} & \psi_{G,i}\end{bmatrix}^{\mathsf T}, \\
u_{G,i}&=\omega_{G,i}\in\mathcal{U}_G,
& \mathcal{U}_G&\triangleq[-\bar{\omega}_G,\bar{\omega}_G], \\
\dot{\mathbf{x}}_{G,i}&=
\begin{bmatrix}
V_G\cos\psi_{G,i}\\
V_G\sin\psi_{G,i}\\
\omega_{G,i}
\end{bmatrix}.
\end{aligned}
\label{eq:vehicle_dynamics}
\end{equation}
Here $\mathbf{p}_{G,i}$ is planar position, $\psi_{G,i}$ is heading, $V_G$ is the guard-team constant forward speed, and $\bar{\omega}_G$ is the maximum turn rate. The corresponding inertial velocity is $\mathbf{v}_{G,i}=V_G[\cos\psi_{G,i},\sin\psi_{G,i}]^{\mathsf T}$. Bandits are defined analogously by replacing $(G,i)$ with $(B,j)$ and using $V_B$, $\bar{\omega}_B$, and $u_{B,j}$. Thus each vehicle state is three-dimensional and each control is a scalar heading rate. Stacking the dynamics over both teams defines $\dot{\mathbf{x}}=f(\mathbf{x},\mathbf{u}_G,\mathbf{u}_B)$. We ignore altitude changes and assume $V_G>V_B$; the rationale for excluding faster-bandit cases is discussed in Sec.~\ref{sec:out_of_scope}.

The full-team and active-roster guard stacks are
\begin{equation}
\begin{aligned}
\mathbf{x}_G(t)&\triangleq
\begin{bmatrix}\mathbf{x}_{G,0}(t)^{\mathsf T}&\cdots&\mathbf{x}_{G,i}(t)^{\mathsf T}&\cdots&\mathbf{x}_{G,N_G-1}(t)^{\mathsf T}\end{bmatrix}^{\mathsf T}, \\
\mathbf{u}_G(t)&\triangleq
\begin{bmatrix}u_{G,0}(t)&\cdots&u_{G,i}(t)&\cdots&u_{G,N_G-1}(t)\end{bmatrix}^{\mathsf T}, \\
\mathbf{x}_{G_A}(t)&\triangleq
\bigl[\mathbf{x}_{G,i}(t)\bigr]_{i\in G_A}, \\
\mathbf{u}_{G_A}(t)&\triangleq
\bigl[u_{G,i}(t)\bigr]_{i\in G_A}.
\end{aligned}
\label{eq:team_state_control_notation}
\end{equation}
with analogous definitions for $\mathbf{x}_B$, $\mathbf{u}_B$, $\mathbf{x}_{B_A}$, and $\mathbf{u}_{B_A}$. Brackets indexed by a set indicate a stack over that set in ascending index order. Thus $\mathbf{x}_G$ and $\mathbf{x}_B$ have fixed dimensions $3N_G$ and $3N_B$, whereas $\mathbf{x}_{G_A}$ and $\mathbf{x}_{B_A}$ are compact active-roster stacks. The joint propagated vehicle state is $\mathbf{x}(t)=[\mathbf{x}_G(t)^{\mathsf T},\mathbf{x}_B(t)^{\mathsf T}]^{\mathsf T}$; the active state and control used by a game solve are $\mathbf{x}_A=[\mathbf{x}_{G_A}^{\mathsf T},\mathbf{x}_{B_A}^{\mathsf T}]^{\mathsf T}$ and $\mathbf{u}_A=[\mathbf{u}_{G_A}^{\mathsf T},\mathbf{u}_{B_A}^{\mathsf T}]^{\mathsf T}$. Superscripts $0:T$ and $0:T-1$ denote planning-horizon state and control sequences; for example, $\mathbf{x}_{G_A}^{0:T}=(\mathbf{x}_{G_A,0},\ldots,\mathbf{x}_{G_A,T})$ and $\mathbf{u}_{G_A}^{0:T-1}=(\mathbf{u}_{G_A,0},\ldots,\mathbf{u}_{G_A,T-1})$.

The HVAs are fixed at positions collected in $\mathbf{P}_L\in\mathbb{R}^{N_L\times2}$ and are scenario context rather than dynamic-state components. Every guard receives the common global observation
\begin{equation}
\mathcal{O}_G\triangleq(\mathbf{x}_G,\mathbf{x}_B,\mathbf{P}_L,\mathbf{V}_G,\mathbf{V}_B,\boldsymbol{\sigma}_G,\boldsymbol{\sigma}_B,\boldsymbol{\sigma}_L,t),
\label{eq:guard_observation_space}
\end{equation}
where $\mathbf{V}_G$ and $\mathbf{V}_B$ are fixed vehicle-speed vectors and the status vectors preserve fixed observation dimensions after entities become inactive. The implemented bandit behavior uses the guard-blind projection $\mathcal{O}_B\triangleq(\mathbf{x}_B,\mathbf{P}_L,\mathbf{V}_B,\boldsymbol{\sigma}_B,\boldsymbol{\sigma}_L,t)$; its actions do not respond to guard state.

\subsection{Objectives and Solution Concepts}

Guards are modeled as a cooperative team with common cost $J_G$, while bandits share cost $J_B$. The detailed terminal-handoff and control-effort costs are specified in Sec.~\ref{sec:defend_mg_gametheory}. Let $\gamma_G:\mathcal{O}_G\rightarrow\mathcal{U}_G^{N_G}$ and $\gamma_B:\mathcal{O}_B\rightarrow\mathcal{U}_B^{N_B}$ denote guard and bandit feedback policies---also called ``strategies'' in game-theory literature~\cite{basar1999dynamic} and colloquially as ``tactics'' in this report. %
A unilateral formulation of the swarm defense problem selects the guard tactics that minimize $J_G$:

\begin{equation}
    \begin{aligned}
        \gamma_G^{\star}
        &=
        \operatorname*{arg\,min}_{\gamma_G}
        J_G\!\left(\gamma_G;\gamma_B\right), \\
        \text{such that}\qquad
        \dot{\mathbf{x}}
        &= f\!\left(\mathbf{x},\gamma_G(\mathcal{O}_G),\gamma_B(\mathcal{O}_B)\right),
        \qquad \mathbf{x}(0)=\mathbf{x}_0 .
    \end{aligned}
    \label{eqn:swarm_defense_optim_problem}
\end{equation}
where $f\!\left(\mathbf{x},\gamma_G(\mathcal{O}_G),\gamma_B(\mathcal{O}_B)\right)$ describes the physical dynamics of the drone swarm.

Equation~\ref{eqn:swarm_defense_optim_problem} is the unilateral formulation used by the baseline tactics in Sec.~\ref{sec:baseline_tactics}: it optimizes the guard policy for a prescribed bandit policy. The differential-game formulation instead solves for mutually responsive strategies.

\begin{equation}
    \begin{aligned}
        J_G\!\left(\gamma_G^{\star},\gamma_B^{\star}\right)
        &\leq
        J_G\!\left(\gamma_G,\gamma_B^{\star}\right), \\
        J_B\!\left(\gamma_G^{\star},\gamma_B^{\star}\right)
        &\leq
        J_B\!\left(\gamma_G^{\star},\gamma_B\right),
    \end{aligned}
    \label{eqn:swarm_defense_nash_equilibrium}
\end{equation}

Equation~\ref{eqn:swarm_defense_nash_equilibrium} defines a \emph{Nash equilibrium}, from which neither team benefits by unilaterally changing its strategy~\cite[Chapter~6]{basar1999dynamic}. The planned bandit strategy is an adversarial model rather than a prediction of the simulated guard-blind bandit behavior. Receding-horizon replanning lets the guard team respond as observed behavior departs from that model.

\section{Guard/Defender Tactics}\label{sec:defend_tactics}

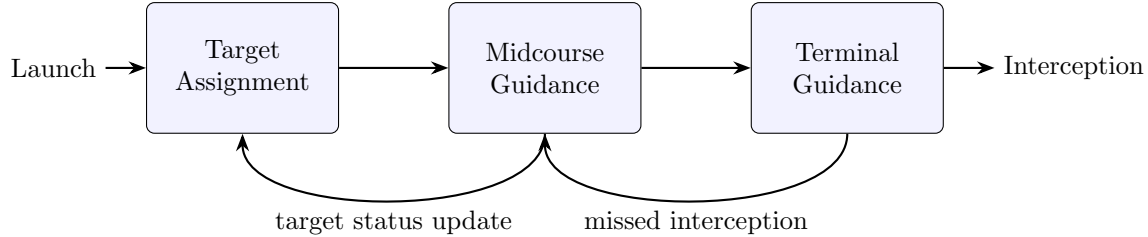
\begin{figure}[t]
\centering
\begin{tikzpicture}[
    stage/.style={draw, rounded corners=3pt, minimum width=1.0in,
                  minimum height=0.68in, align=center, fill=blue!5},
    flow/.style={->, thick, >=Stealth}
]
    \node[stage] (assignment) at (0,0) {Target\\Assignment};
    \node[stage] (midcourse) at (4.0,0) {Midcourse\\Guidance};
    \node[stage] (terminal) at (8.0,0) {Terminal\\Guidance};
    \node (launch) at (-2.5,0) {Launch};
    \node (interception) at (11.0,0) {Interception};

    \draw[flow] (launch.east) -- (assignment.west);
    \draw[flow] (assignment) -- (midcourse);
    \draw[flow] (midcourse) -- (terminal);
    \draw[flow] (terminal.east) -- (interception.west);
    \draw[flow] (terminal.south)
        to[out=-90, in=-90, looseness=0.75]
        node[midway, below] {missed interception}
        (midcourse.south);

    \draw[flow] (midcourse.south)
        to[out=-90, in=-90, looseness=0.75]
        node[midway, below] {target status update}
        (assignment.south);
\end{tikzpicture}
\caption{High-level control flow for guard/defender tactics. 
}
\label{fig:guard_tactics_flow}
\end{figure}

In this section we develop 3-phase tactics used by the team of guard drones; the phases being target assignment, midcourse guidance (MG), and terminal guidance (TG). Fig.~\ref{fig:guard_tactics_flow} provides an illustrative diagram of the control flow between these phases. %
Generally speaking, the target assignment phase is responsible for assigning to each guard drone a bandit that it is nominally intended to intercept. This phase is generally executed by a centralized planner---mimicking how a command and control node in the real-world would be responsible for sensing incoming threats and tasking interceptors to meet those threats---and may be run multiple times throughout the engagement, leading to dynamic reassignment of targets (Sec.~\ref{sec:defend_target_assignment}). %
The midcourse guidance phase is generally responsible for planning the flyout trajectory of the guards from their current state to a state that best establishes the conditions for an interception. This phase receives much of the attention in the subsequent sections as we develop novel tactics based upon differential game theory (DG) that blend the midcourse guidance and target assignment phases. %
Terminal guidance is generally responsible for control of a guard drone in the final moments before interception of a bandit. We employ the widely-used, highly-robust technique of proportional navigation (PN) for all of our terminal guidance algorithms; see Sec.~\ref{sec:defend_tg} for more discussion of this guidance law.

Not only do we develop the novel game-theoretic tactics (Sec.~\ref{sec:dg_tactics}) that are the subject of investigation in this report, but also a set of baseline tactics (Sec.~\ref{sec:baseline_tactics}) against which the differential game tactics are compared. %
All guard tactics, both game-theoretic and baselines, are based upon the same 3-phase architecture given in Fig.~\ref{fig:guard_tactics_flow}. %
They differ in the algorithms that underpin the various phases; most importantly, the midcourse guidance phase. %

\subsection{Baseline Tactics}
\label{sec:baseline_tactics}

To evaluate the effectiveness of the differential game tactics described in Sec.~\ref{sec:dg_tactics}, it is essential that operationally realistic baseline tactics are developed for comparison. %
Two baseline tactics are developed: nearest-bandit (NB) and coverage-aware (CA). %
Both of these tactics are based upon an optimization model of the swarm defense problem---i.e. Eq.~\ref{eqn:swarm_defense_optim_problem}, in contrast to the game-theoretic model of Eq.~\eqref{eqn:swarm_defense_nash_equilibrium}---and share the same general solution algorithm, presented in Alg.~\ref{alg:baseline_tactics}. %
This algorithm is an instantiation of the 3-phase tactics presented in Fig.~\ref{fig:guard_tactics_flow}. %

The nearest-bandit (NB) and coverage-aware (CA) tactics differ only in their target assignment phase (Alg.~\ref{alg:baseline_tactics}, Line~\ref{alg:baseline_tactics:target_assignment}). %
As the name implies, the nearest-bandit tactics simply assign to each guard drone the bandit nearest them for interception. %

The coverage-aware tactics use a more sophisticated target assignment subroutine that balances between \emph{ease} of interception (e.g. distance from target at assignment) and coverage of all bandits (i.e. incentivizes at least one guard to be assigned to each bandit). This target assignment algorithm is based upon the discounted-redundancy linear assignment optimization described in Sec.~\ref{sec:defend_target_assignment}. %

Algorithm~\ref{alg:baseline_tactics} works in an iterative, receding-horizon fashion. %
At the start of each iteration, a centralized control node checks if there are still active guard and bandit drones. %
Line~\ref{alg:baseline_tactics:target_assignment} represents the target assignment phase previously described. %
The rest of the algorithm describes how the midcourse guidance (Line~\ref{alg:baseline_tactics:heading_error}) and terminal guidance (Line~\ref{alg:baseline_tactics:pn_control}) phases are employed in a receding-horizon fashion with updates to the target assignments every $T_{\mathrm{replan}}$ seconds (Line~\ref{alg:baseline_tactics:replan_loop}) or when guards or bandits become inactive due to interceptions (Line~\ref{alg:baseline_tactics:gb_update_break}). %
The $\mathsf{HeadingErrorGuidance}$ midcourse guidance subroutine is described in Sec.~\ref{sec:heading_error}. %
The $\mathsf{ExecuteControls}$ subroutine (Line~\ref{alg:baseline_tactics:execute_controls}) is where the selected controls are employed and the state of the drone swarms are updated.

\begin{algorithm}
\caption{Swarm Defense Baseline Tactics}
\label{alg:baseline_tactics}
\begin{algorithmic}[1]
\While{$G_A \neq \varnothing$ \textbf{and} $B_A \neq \varnothing$}
    \State $\bigl(p_{ij}\bigr) \gets \mathsf{AssignTargets}(G_A, B_A)$
    \label{alg:baseline_tactics:target_assignment}
    \For{$t \gets 0$ \textbf{to} $T_{\mathrm{replan}}$} 
    \label{alg:baseline_tactics:replan_loop}
        \ForAll{$i \in G_A$}
            \State $j \gets \operatorname*{arg\,max}_{j'\in B_A} p_{ij'}$
            \If{$\mathsf{TerminalHandoffAllowed}(\mathbf{x}_{G,i}, \mathbf{x}_{B,j})$}
            \label{alg:baseline_tactics:tg_handoff}
                \State $u_{G,i} \gets \mathsf{ProportionalNavigation}(\mathbf{x}_{G,i}, \mathbf{x}_{B,j})$
                \label{alg:baseline_tactics:pn_control}
            \Else
                \State $u_{G,i} \gets \mathsf{HeadingErrorGuidance}(\mathbf{x}_{G,i}, \mathbf{x}_{B,j})$
                \label{alg:baseline_tactics:heading_error}
            \EndIf
        \EndFor
        \State $(\mathbf{x}_G, \mathbf{x}_B, G_A, B_A) \gets \mathsf{ExecuteControls}(\mathbf{u}_G)$
        \label{alg:baseline_tactics:execute_controls}
        \If{$G_A$ or $B_A$ has changed}
        \label{alg:baseline_tactics:gb_update_break}
            \State \textbf{break}
        \EndIf
    \EndFor
\EndWhile
\end{algorithmic}
\end{algorithm}

\subsubsection{Heading Error Midcourse Guidance}
\label{sec:heading_error}

With the $\mathsf{HeadingErrorGuidance}$ controller, the guard seeks to align its heading angle with the line-of-sight (LOS) to the assigned bandit as described in the following equations:
\begin{equation}
    \begin{aligned}
        \psi_{\mathrm{LOS},j/i}
        &= \operatorname{atan2}\!\left(
            p_{y,B,j} - p_{y,G,i},\,
            p_{x,B,j} - p_{x,G,i}
        \right), \\
        e_{\psi,i}
        &= \operatorname{atan2}\!\left(
            \sin\!\left(\psi_{\mathrm{LOS},j/i} - \psi_{G,i}\right),\,
            \cos\!\left(\psi_{\mathrm{LOS},j/i} - \psi_{G,i}\right)
        \right), \\
        u_{G,i} &= k_g e_{\psi,i}.
    \end{aligned}
\end{equation}
where $u_{G,i}$ is the commanded turn rate of guard $i$, $k_g$ is a positive heading-error gain, $\psi_{\mathrm{LOS},j/i}$ is the line-of-sight angle from guard $i$ to bandit $j$, and $e_{\psi,i}$ is the heading error wrapped to $[-\pi,\pi]$. The state-component notation is defined in Sec.~\ref{sec:vehicle_dynamics}.

\subsection{Differential Game Tactics}\label{sec:dg_tactics}

Algorithm~\ref{alg:dg_tactics} provides pseudocode for implementing the differential game (DG) tactics. %
Similar to Alg.~\ref{alg:baseline_tactics}, the DG tactics represent an instantiation of the 3-phase tactics presented in Fig.~\ref{fig:guard_tactics_flow}. %
Algorithm~\ref{alg:dg_tactics} also works in an iterative, receding-horizon fashion. %
At the start of each iteration, a centralized control node checks if there are still active guard and bandit drones. %
Line~\ref{alg:dg_tactics:target_assignment} represents the target assignment phase, which for the DG tactics, is actually a soft-assignment; i.e. a \emph{first-guess} at which bandits each guard should pursue. Sec.~\ref{sec:defend_target_assignment} describes the $\mathsf{AssignTargets}$ subroutine for this soft-assignment. %

Lines~\ref{alg:dg_tactics:form_ref_traj}--\ref{alg:dg_tactics:solve_feedback_nash} represent the game-theoretic components of the midcourse guidance law which are described more thoroughly in Sec.~\ref{sec:defend_mg_gametheory}. These components are based upon the Nash equilibrium (Eq.~\ref{eqn:swarm_defense_nash_equilibrium}) of a linear-quadratic approximation of the swarm defense differential game. %
The rest of the algorithm describes how the differential game strategy is employed in a receding-horizon fashion with updates to the target soft-assignments every $T_{\mathrm{replan}}$ seconds (Line~\ref{alg:dg_tactics:replan_loop}) or when guards or bandits become inactive due to interceptions (Line~\ref{alg:dg_tactics:gb_update_break}). Lines~\ref{alg:dg_tactics:tg_handoff} and \ref{alg:dg_tactics:pn_control} represent the terminal guidance handoff to the proportional navigation controller described in Sec.~\ref{sec:defend_tg}.

\begin{algorithm}
\caption{Swarm Defense Differential Game Tactics}
\label{alg:dg_tactics}
\begin{algorithmic}[1]
\While{$G_A \neq \varnothing$ \textbf{and} $B_A \neq \varnothing$}
    \State $\bigl(p_{ij}\bigr) \gets \mathsf{AssignTargets}(G_A, B_A)$
    \label{alg:dg_tactics:target_assignment}
    \State $(\bar{\mathbf{x}}_A^{0:T},\bar{\mathbf{u}}_A^{0:T-1}) \gets \mathsf{FormReferenceTrajectories}\bigl((p_{ij}),\mathbf{x}_{G_A},\mathbf{x}_{B_A}\bigr)$
    \label{alg:dg_tactics:form_ref_traj}
    \State $(J_G,J_B) \gets \mathsf{DefineGameCosts}(G_A,B_A)$
    \State $\mathit{lqgame} \gets \mathsf{ApproxLQGame}(\bar{\mathbf{x}}_A^{0:T},\bar{\mathbf{u}}_A^{0:T-1},J_G,J_B)$
    \State $\gamma_G \gets \mathsf{SolveFeedbackNash}(\mathit{lqgame})$
    \label{alg:dg_tactics:solve_feedback_nash}
    \For{$t \gets 0$ \textbf{to} $T_{\mathrm{replan}}$} \label{alg:dg_tactics:replan_loop}
        \ForAll{$i \in G_A$}
            \If{$\mathsf{TerminalHandoffAllowed}(i, \mathbf{x}_G, \mathbf{x}_B)$} \label{alg:dg_tactics:tg_handoff}
                \State $u_{G,i} \gets \mathsf{ProportionalNavigation}(i, \mathbf{x}_G, \mathbf{x}_B)$ \label{alg:dg_tactics:pn_control}
            \Else
                \State $u_{G,i} \gets \gamma_{G,i}(t,\mathbf{x}_{G_A},\mathbf{x}_{B_A})$
            \EndIf
        \EndFor
        \State $(\mathbf{x}_G, \mathbf{x}_B, G_A, B_A) \gets \mathsf{ExecuteControls}(\mathbf{u}_G)$
        \If{$G_A$ or $B_A$ has changed} \label{alg:dg_tactics:gb_update_break}
            \State \textbf{break}
        \EndIf
    \EndFor
\EndWhile
\end{algorithmic}
\end{algorithm}

\subsubsection{Dynamic Target Soft-Assignment}\label{sec:defend_target_assignment}

In the $\mathsf{AssignTargets}$ subroutine of Alg.~\ref{alg:dg_tactics} (Line~\ref{alg:dg_tactics:target_assignment}), a shared coordinator, akin to a centralized command and control node, provides \emph{soft-assignments} between active guards and incoming bandits to be intercepted. %
The assignment is ``soft'' in the sense that it is only an initial guess as to which bandit a guard will intercept; however, the midcourse guidance phase (Sec.~\ref{sec:defend_mg_gametheory}) may ultimately result in a different interception.

Assigning to each guard a bandit to be pursued requires an algorithm that accounts for the relative \emph{difficulty} of all possible guard-bandit pairings and balances this against the desire to intercept each and every bandit, with the constraint that each guard may only target one bandit at a time. %
In order to handle all cases of relative guard/bandit swarm sizes, the soft-assignment algorithm must be---in general---a many-to-one, potentially non-surjective function; i.e. multiple elements of the guard set may target the same bandit in the bandit set, and some bandits may have no guard assigned to them. %

Using the total and active-roster notation of Sec.~\ref{sec:teams_state_information}, let $p_{ij} \in \{0,1\}$ indicate whether guard $i\in G_A$ is assigned to bandit $j\in B_A$. Let the number of guards assigned to bandit $j$ be
\begin{equation}
    P_j = \sum_{i \in G_A} p_{ij}.
\end{equation}

To model the assignment problem, we assume that there is both a \emph{cost}, $c^{\mathrm{TA}}_{ij} \geq 0$, associated with pairing $p_{ij}$, as well as a reward, $v_j \geq 0$, for having had \emph{some} guard assigned to bandit $j$ (i.e. the assignment reward is agnostic to which guard is assigned). %
We then apply a redundancy discount factor, $\beta \in \left[0, 1\right]$, that reduces the marginal reward of assigning more than one guard to a bandit. Therefore, we have the total payoff of assignments to bandit $j$ given as
\begin{equation}
    V_j(P_j) = v_j \sum_{k=0}^{P_j-1} \beta^k,
    \qquad V_j(0) = 0.
\end{equation}
Thus, the first assigned guard contributes reward $v_j$, the second adds
$\beta v_j$, and each additional guard has weakly smaller marginal reward.

Therefore, we can define the target (soft) assignment problem as
\begin{align}
    \max_{p_{ij}} \quad &
    \sum_{j \in B_A} V_j(P_j)
    - \sum_{i \in G_A} \sum_{j \in B_A} c^{\mathrm{TA}}_{ij} p_{ij},
    \\
    \text{subject to} \quad &
    \sum_{j \in B_A} p_{ij} = 1, \qquad \forall i \in G_A,
    \\
    & p_{ij} \in \{0,1\}, \qquad \forall i \in G_A,\; j \in B_A.
\end{align}
Each guard must be assigned to one-and-only-one bandit, while a bandit may receive zero,
one, or multiple guards. The case $\beta = 0$ values only first coverage of
each bandit; values $0 < \beta < 1$ reward redundancy with diminishing returns;
and $\beta = 1$ assigns full value to every additional guard. 

In our experiments, the cost, bandit reward, and redundancy discount are
\begin{equation}
    c^{\mathrm{TA}}_{ij}
    = w_{\mathrm{TA}}
    \frac{\lVert \mathbf{p}_{B,j}-\mathbf{p}_{G,i}\rVert}
    {\displaystyle\max_{i'\in G_A,\,j'\in B_A}
    \lVert \mathbf{p}_{B,j'}-\mathbf{p}_{G,i'}\rVert},
    \qquad
    v_j = 1,
    \qquad
    \beta = 0.1,
    \qquad
    w_{\mathrm{TA}}=0.25,
    \label{eq:target_assignment_parameters}
\end{equation}
where $c^{\mathrm{TA}}_{ij}=0$ when the normalizing maximum is zero.

The discounted-redundancy objective is solved by expanding each active bandit into $N_{G_A}$ virtual assignment slots with successively discounted marginal rewards, then applying a rectangular linear-assignment solver. The selected slot for each guard maps directly to its associated physical bandit.
The full linear reformulation is given in Appendix~\ref{app:discounted_assignment}.

\subsubsection{Game Theoretic Midcourse Guidance}
\label{sec:defend_mg_gametheory}

The differential-game (DG) midcourse guidance in lines~\ref{alg:dg_tactics:form_ref_traj}--\ref{alg:dg_tactics:solve_feedback_nash} of Alg.~\ref{alg:dg_tactics} repeatedly constructs and solves a two-player, finite-horizon approximation of the swarm-defense game whose high-level definition is given in Eq.~\eqref{eqn:swarm_defense_nash_equilibrium}. The bandit player is an adversarial planning model, rather than a prediction of the simulated bandits' guard-blind behavior (Sec.~\ref{sec:teams_state_information}).

At each replan, $(p_{ij})$ supplies one provisional bandit target for each active guard. These assignments seed the nominal active-roster trajectories $(\bar{\mathbf{x}}_A^{0:T},\bar{\mathbf{u}}_A^{0:T-1})$: guards use heading-error guidance toward their provisionally assigned bandits, while bandits use zero nominal turn rate. Propagation under the nonlinear Dubins dynamics produces a dynamically feasible operating trajectory. The provisional assignment is used only for this initialization; the terminal objective---later summarized in Eq.~\eqref{eq:guard_terminal_handoff_cost}---evaluates every pair in $G_A\times B_A$ and can therefore favor a different pairing.

At its core, the DG guidance law encodes the guard team's desire for optimal conditions \emph{at the end of midcourse guidance} to handoff to terminal guidance. %
More specifically, guards favor terminal-handoff pairs that are close in distance and require little PN maneuvering (Sec.~\ref{sec:defend_tg}). %
Therefore, at the terminal handoff time $t_f$, each active guard--bandit pair is evaluated by the differentiable terminal-handoff kernel $c_{ij}^{\mathrm{TH}}$. Let $\mathbf{r}_{j/i}=\mathbf{p}_{B,j}-\mathbf{p}_{G,i}$ and $\mathbf{v}_{j/i}=\mathbf{v}_{B,j}-\mathbf{v}_{G,i}$ denote the relative position and velocity from guard $i$ to bandit $j$.
\begin{equation}
    \begin{aligned}
        R_{\epsilon,ij}
        &= \sqrt{\mathbf{r}_{j/i}^{\mathsf{T}}\mathbf{r}_{j/i}
            + \epsilon_R^2}, \\
        d_{\epsilon,ij} &= R_{\epsilon,ij}-\epsilon_R, \\
        V_{c,ij}
        &= -\frac{\mathbf{r}_{j/i}^{\mathsf{T}}\mathbf{v}_{j/i}}
        {R_{\epsilon,ij}}, \\
        \dot{\lambda}_{ij}
        &= \frac{r_{x,j/i}v_{y,j/i}-r_{y,j/i}v_{x,j/i}}
        {R_{\epsilon,ij}^{2}}, \\
        a_{\mathrm{PN},ij}
        &= \alpha_G V_{c,ij}\dot{\lambda}_{ij}, \\
        c_{ij}^{\mathrm{TH}}
        &= w_{gb}d_{\epsilon,ij}
        +w_{\mathrm{PN}}\left(
            \sqrt{a_{\mathrm{PN},ij}^{2}+\epsilon_a^2}-\epsilon_a
        \right).
        \label{eq:terminal_handoff_cost}
    \end{aligned}
\end{equation}
Here $\alpha_G$ is the guard PN navigation constant (typically set to $3.0$; see Sec.~\ref{sec:defend_tg}) and $V_{c,ij}>0$ denotes a closing engagement. The two positive regularization parameters make the range and acceleration-magnitude terms differentiable at zero range and zero lateral acceleration; away from these geometries, they approach $\lVert\mathbf{r}_{j/i}\rVert$ and $\lvert a_{\mathrm{PN},ij}\rvert$, respectively. %

The guard and bandit swarms are each controlled by one team coordinator. Rather than fixing the provisional assignment in the terminal objective, the guard team takes a differentiable soft minimum (i.e. LogSumExp) over every active bandit for each active guard:

\begin{equation}
    g_{f,G}^{\mathrm{TH}}
    =
    \begin{cases}
        \displaystyle
        \sum_{i\in G_A}
        \left[-\tau\log\!\left(
            \sum_{j\in B_A}
            \exp\!\left(-\frac{c_{ij}^{\mathrm{TH}}}{\tau}\right)
        \right)\right],
        & B_A\neq\varnothing, \\[1.0em]
        0, & B_A=\varnothing,
    \end{cases}
    \label{eq:guard_terminal_handoff_cost}
\end{equation}
where $\tau>0$ is the soft-assignment temperature. As $\tau$ approaches zero, the LogSumExp soft minimum approaches the hard minimum, but becomes less numerically well conditioned under differentiation. The terminal bandit objective is its negative, $g_{f,B}^{\mathrm{TH}}=-g_{f,G}^{\mathrm{TH}}$.

Turn-rate limits are represented in the game by normalized quadratic effort penalties:

\begin{equation}
    \begin{aligned}
        g_{t,G}(\mathbf{u}_{G_A})
        &= w_{u,G}\sum_{i\in G_A}
        \left(\frac{u_{G,i}}{\bar{\omega}_G}\right)^2, \\
        g_{t,B}(\mathbf{u}_{B_A})
        &= w_{u,B}\sum_{j\in B_A}
        \left(\frac{u_{B,j}}{\bar{\omega}_B}\right)^2.
        \label{eq:running_control_costs}
    \end{aligned}
\end{equation}
where $\bar{\omega}_G$ and $\bar{\omega}_B$ are the maximum turn rates and $w_{u,G}$ and $w_{u,B}$ are effort weights. The complete team costs are

\begin{equation}
    \begin{aligned}
        J_G
        &= g_{f,G}^{\mathrm{TH}}\bigl(\mathbf{x}_A(t_f)\bigr)
        + \int_{t_0}^{t_f}g_{t,G}\bigl(\mathbf{u}_{G_A}(t)\bigr)\,dt, \\
        J_B
        &= g_{f,B}^{\mathrm{TH}}\bigl(\mathbf{x}_A(t_f)\bigr)
        + \int_{t_0}^{t_f}g_{t,B}\bigl(\mathbf{u}_{B_A}(t)\bigr)\,dt,
        \qquad
        g_{f,B}^{\mathrm{TH}}=-g_{f,G}^{\mathrm{TH}}.
        \label{eq:team_game_costs}
    \end{aligned}
\end{equation}

The nonlinear dynamics and costs in Eqs.~\eqref{eq:vehicle_dynamics} and \eqref{eq:team_game_costs} are locally approximated about the nominal trajectory. With $\delta\mathbf{x}_k=\mathbf{x}_{A,k}-\bar{\mathbf{x}}_{A,k}$ and $\delta\mathbf{u}_k=\mathbf{u}_{A,k}-\bar{\mathbf{u}}_{A,k}$, the discrete-time dynamics and each team cost are approximated as
\begin{equation}
\begin{aligned}
\delta\mathbf{x}_{k+1} &\approx A_k\delta\mathbf{x}_k+B_k\delta\mathbf{u}_k, \\
J_q &\approx J_q(\bar{\mathbf{x}}_A^{0:T},\bar{\mathbf{u}}_A^{0:T-1})
+\mathbf{h}_q^{\mathsf T}\delta\mathbf{z}
+\tfrac{1}{2}\delta\mathbf{z}^{\mathsf T}H_q\delta\mathbf{z},
\qquad q\in\{G,B\},
\end{aligned}
\label{eq:local_lq_approximation}
\end{equation}
where $A_k$ and $B_k$ are the dynamics Jacobians and $\delta\mathbf{z}$ stacks the state and control deviations over the planning horizon. The gradients and Hessians defining $\mathbf{h}_q$ and $H_q$ are obtained by automatic differentiation. These linear dynamics and quadratic costs form an LQ game, whose feedback Nash equilibrium is computed by Riccati recursion~\cite{basar1999dynamic,fridovichkeil2020efficient} to obtain the guard policy $\gamma_G$. %
In practice, the PYDGENS library~\cite{allen2026pydgens} is used for both the LQ approximation ($\mathrm{ApproxLQGame}$) and Riccati recursion ($\mathrm{SolveFeedbackNash}$).

The guards execute $\gamma_G$ during midcourse guidance until a guard transitions to PN terminal guidance. Only a short prefix of the planned policy is used: the cycle repeats after $T_{\mathrm{replan}}$ or whenever $G_A$ or $B_A$ changes, rebuilding the assignment, nominal trajectory, and local game from the new state.

\subsubsection{Terminal Guidance}
\label{sec:defend_tg}

Proportional navigation (PN) is a terminal homing law that commands lateral acceleration in proportion to the line-of-sight (LOS) angular rate. It is widely used because it is comparatively simple while directly responding to the relative geometry that determines miss distance \cite{palumbo2010homing}.

For guard $i\in G_A$ and bandit $j\in B_A$, use the relative position and velocity introduced for the terminal-handoff cost in Eq.~\eqref{eq:terminal_handoff_cost}: $\mathbf{r}_{j/i}=\mathbf{p}_{B,j}-\mathbf{p}_{G,i}$ and $\mathbf{v}_{j/i}=\mathbf{v}_{B,j}-\mathbf{v}_{G,i}$. Let $R_{ij}=\lVert\mathbf{r}_{j/i}\rVert$ and $\lambda_{ij}=\operatorname{atan2}(r_{y,j/i},r_{x,j/i})$. The planar LOS rate and closing speed are
\begin{equation}
    \dot{\lambda}_{ij}=\frac{r_{x,j/i}v_{y,j/i}-r_{y,j/i}v_{x,j/i}}{R_{ij}^2},
    \qquad
    V_{c,ij}=-\frac{\mathbf{r}_{j/i}^{\mathsf{T}}\mathbf{v}_{j/i}}{R_{ij}},
\end{equation}
where $V_{c,ij}>0$ denotes decreasing range. The scalar PN normal-acceleration command for guard $G_i$ against bandit $B_j$ is
\begin{equation}
    a_{\mathrm{PN},ij} = \alpha_G V_{c,ij}\dot{\lambda}_{ij},
    \label{eq:planar_pn}
\end{equation}
where $\alpha_G>0$ is the guard's dimensionless navigation constant. The sign of $a_{\mathrm{PN},ij}$ selects the turn direction. For a fixed-speed planar Dubins guard with speed $V_G$, this lateral-acceleration command corresponds to the heading-rate command $\omega_{\mathrm{PN},ij}=a_{\mathrm{PN},ij}/V_G$.

The scalar law in Eq.~\eqref{eq:planar_pn} does not by itself specify the direction in which guard $G_i$ realizes the commanded acceleration. In \emph{true PN}, the commanded acceleration is normal to the LOS; in \emph{pure PN}, it is normal to the guard velocity (equivalently, normal to its body axis in this planar fixed-speed model). The terminal guard policy implements pure PN by converting the scalar command to a Dubins turn rate. The differential-game cost in Eq.~\eqref{eq:terminal_handoff_cost} uses the magnitude of that same scalar command only as a measure of terminal-handoff maneuvering demand; it does not propagate a PN controller during the midcourse game solve.

\section{Simulation Environment}
\label{sec:sim_env}

The guard tactics (Sec.~\ref{sec:defend_tactics}) are evaluated within a simulation environment built in Python using the PettingZoo API to model multi-agent decision processes~\cite{terry2021pettingzoo}. The linear-quadratic game in the DG tactics (Alg.~\ref{alg:dg_tactics}, Line~\ref{alg:dg_tactics:solve_feedback_nash}) is solved with the PYDGENS library~\cite{allen2026pydgens}. %
The system dynamics (Eq.~\eqref{eq:vehicle_dynamics}) are integrated over fixed time steps, typically of 1.0 seconds. %
The information and action structure of the simulated teams is defined in Sec.~\ref{sec:teams_state_information}; the remainder of this section describes its implementation and engagement-specific assumptions. Section~\ref{sec:intercept_func} describes how interceptions of bandits and assets are adjudicated. %
In contrast to the guard team which is modeled as an \emph{agent}---i.e. whose tactics are separable from the simulation environment---the bandit team is modeled as a \emph{bot} whose tactics are part of the environment. Section~\ref{sec:intrude_tactics} describes two different scenarios for the simulation based upon differing bandit behaviors: non-evasive bandits (B1) and evasive bandits (B2).
Section~\ref{sec:exp_design} then describes how the simulation environment is used in a set of Monte Carlo experiments, along with the performance metrics and Bayesian statistics used for evaluation of tactics performance.

\subsection{Interception Adjudication}
\label{sec:intercept_func}

The determination of whether a simulated guard intercepts a bandit and whether a simulated bandit reaches a HVA is key to the evaluation of the ``success'' of either team. %
When a guard successfully intercepts a bandit, both vehicles are eliminated; the bandit then poses no threat to assets and the guard is not able to intercept any further bandits. 
However, due to large simulation time steps (e.g. 1.0 seconds) used to accelerate computation in large Monte Carlo campaigns, we cannot simply implement a narrow interception range as it would often be passed through, unrecorded, from one time step to the next. 
Instead, we use a dynamically-scaled \emph{effective} interception range (Eq.~\eqref{eq:gb_interception}) that increases in size when guard and bandit headings are aligned with a small relative velocity. %
This tends to bias the interception adjudication toward approach-from-behind configurations, which is a common technique for real-world interceptors.

For each active guard--bandit pair $(i,j)\in G_A\times B_A$, reuse the relative position $\mathbf{r}_{j/i}$ and velocity $\mathbf{v}_{j/i}$ defined for the terminal-handoff kernel in Eq.~\eqref{eq:terminal_handoff_cost}. Let $R_{ij}=\lVert\mathbf{r}_{j/i}\rVert$ and $\hat{\boldsymbol{\ell}}_{j/i}=\mathbf{r}_{j/i}/R_{ij}$ denote the corresponding range and guard-to-bandit line-of-sight unit vector. The environment declares an interception when
\begin{equation}
    \begin{aligned}
        \mathsf{intercept}_{ij}
        &\Longleftrightarrow
        R_{ij} \leq R_{\mathrm{eff},ij}, \\
        R_{\mathrm{eff},ij}
        &= R_{\mathrm{gb}}
        + \left(R_{\mathrm{adj}} - R_{\mathrm{gb}}\right)
        \underbrace{\operatorname{clip}_{[0,1]}\!\left(
            \frac{-\hat{\boldsymbol{\ell}}_{j/i}^{\mathsf{T}}\mathbf{v}_{j/i}}
            {\lVert\mathbf{v}_{j/i}\rVert}
        \right)}_{\text{closing alignment}}
        \underbrace{\operatorname{clip}_{[0,1]}\!\left(
            1 - \frac{\lVert\mathbf{v}_{j/i}\rVert}{V_{\mathrm{rel}}}
        \right)}_{\text{relative-speed factor}} .
    \end{aligned}
    \label{eq:gb_interception}
\end{equation}
Here $R_{\mathrm{gb}}$ is the physical guard--bandit interception radius, $R_{\mathrm{adj}}$ is the maximum adjudication range, and $V_{\mathrm{rel}}$ is the relative-speed scale. Thus, the envelope expands beyond the physical radius only for a closing, nearly collinear, low-relative-speed encounter; it reduces to $R_{\mathrm{gb}}$ for non-closing geometry or relative speed at or above $V_{\mathrm{rel}}$. The implementation assigns zero closing alignment when either range or relative speed is zero.

\subsection{Bandit/Intruder Tactics}
\label{sec:intrude_tactics}

The swarm of bandit drones also employs 3-phase tactics, as depicted in Fig.~\ref{fig:guard_tactics_flow}, with target assignment, midcourse guidance, and terminal guidance phases. %
The bandit tactics differ from the baseline guard tactics (Sec.~\ref{sec:baseline_tactics}) in that they employ static target assignment (i.e. HVA targets are assigned to each bandit at launch and do not change throughout the engagement), and that they cannot observe---and, therefore, are not reactive to---the state of the guard drones (see Sec.~\ref{sec:teams_state_information}).

Two different behavior models for the bandits are studied. First is that of \emph{non-evasive} midcourse guidance---labeled simply as Behavior 1 or B1---whereby bandits navigate to their assigned HVA targets using heading-error midcourse guidance described in Sec.~\ref{sec:heading_error}. The second behavior---labeled Behavior 2, B2---uses sporadic \emph{evasive} maneuvers during midcourse guidance.

In the B2 version of the simulation environment, each bandit receives an immutable, open-loop ``jink'' schedule at initialization. 
The schedule is constructed based upon the bandit's initial time-to-go to its fixed target, assuming a straight-line trajectory; jinks are placed in non-overlapping intervals spanning the middle portion of that nominal engagement. %
For each scheduled jink, the start time and turn direction are sampled, while the turn-rate magnitude and jink duration are pre-specified parameters of the environment. %
During the jink interval, the nominal heading-error command is replaced with the sampled turn-rate command, temporarily deviating the bandit from its nominal target track. %
At the end of the interval, the bandit resumes the heading-error midcourse guidance control described in Sec.~\ref{sec:heading_error}.

It is worth noting that, in spite of the use of game-theoretic techniques described in Sec.~\ref{sec:dg_tactics} that compute an equilibrium strategy for the bandit swarm, this strategy is not actually in use by the bandits.

\section{Experiment Design}
\label{sec:exp_design}

In order to evaluate the effectiveness of the differential game (DG) tactics (Sec.~\ref{sec:dg_tactics}) compared to baseline tactics (Sec.~\ref{sec:baseline_tactics}), we use the simulation environment (Sec.~\ref{sec:sim_env}) to conduct a Monte Carlo campaign of randomized trials, measure performance metrics, and perform Bayesian analysis of the performance metrics to estimate the probability that differential game tactics outperform the baselines.

\subsection{Monte Carlo Trials}
\label{sec:monte_carlo_setup}

Two separate Monte Carlo campaigns are conducted: one with the non-evasive bandit behaviors (B1) and one with evasive bandits (B2). %
Table~\ref{tab:trial_parameters} summarizes the parameters that are randomized between Monte Carlo trials.
Unless otherwise indicated, numerical intervals in Table~\ref{tab:trial_parameters} are uniform sampling ranges; initial positions are sampled independently and uniformly within their stated rectangles. The B2 sampler shares every B1 parameter and adds the open-loop jink (i.e. evasive maneuver) parameters marked with an asterisk.

The same random number seed for randomizing trial parameters is used across the coverage-aware (CA) and differential game (DG) experiment campaigns. %
Therefore, the CA and DG tactics are evaluated with trials of identical parameters. %
This enables a more direct comparison of Bayesian posterior distributions of performance metrics; discussed in the next section.

\begin{table}[t]
\centering
\caption{Monte Carlo trial parameters for non-evasive (B1) and evasive (B2) bandit scenarios. Parameters marked
with $^{*}$ apply only to the B2 evasive-bandit scenario.}
\label{tab:trial_parameters}
\small
\renewcommand{\arraystretch}{1.15}
\begin{tabular}{p{1.12in} p{2.30in} p{1.55in} p{0.55in}}
\hline
\textbf{Category} & \textbf{Parameter} & \textbf{Value or sampling range} & \textbf{Scenario} \\
\hline
Campaign & Integration time step, maximum steps & $1~\mathrm{s}$, $500$ ($500~\mathrm{s}$ maximum duration) & B1, B2 \\
\hline
Team size & Assets $N_L$ & $\operatorname{UnifInt}\{4,\ldots,8\}$ & B1, B2 \\
& Bandits $N_B$ & $\operatorname{UnifInt}\{8,\ldots,12\}$ & B1, B2 \\
& Guards $N_G$ & $\operatorname{UnifInt}\{\max(10,N_B),\ldots,16\}$ & B1, B2 \\
\hline
Vehicle & Guard speed & $[70,80]~\mathrm{m/s}$ & B1, B2 \\
& Bandit speed & $[45,55]~\mathrm{m/s}$ & B1, B2 \\
& Guard maximum turn rate & $[0.8,1.2]~\mathrm{rad/s}$ & B1, B2 \\
& Bandit maximum turn rate & $[0.45,0.55]~\mathrm{rad/s}$ & B1, B2 \\
\hline
Adjudication & Guard--bandit physical radius & $[2.5,3.5]~\mathrm{m}$ & B1, B2 \\
& Guard--bandit maximum adjudication range & $[95,105]~\mathrm{m}$ & B1, B2 \\
& Guard--bandit relative-speed scale & $[95,105]~\mathrm{m/s}$ & B1, B2 \\
& Bandit--asset physical radius & $[9.9,10.1]~\mathrm{m}$ & B1, B2 \\
& Bandit--asset adjudication range & $[95,105]~\mathrm{m}$ & B1, B2 \\
\hline
Initial state & Guard and asset position & $x,y \in [-5{,}000,5{,}000]~\mathrm{m}$ & B1, B2 \\
& Bandit position & $x \in [-20{,}000,20{,}000]~\mathrm{m}$; $y \in [10{,}000,15{,}000]~\mathrm{m}$ & B1, B2 \\
& Initial-heading noise half-width & $[0.2,0.3]~\mathrm{rad}$ & B1, B2 \\
\hline
Bandit guidance & PN navigation constant & $[2.5,3.5]$ & B1, B2 \\
& Midcourse heading-error gain & $[0.4,0.6]~\mathrm{s}^{-1}$ & B1, B2 \\
& Terminal-handoff range & $[80,120]~\mathrm{m}$ & B1, B2 \\
& Terminal LOS-error threshold & $[\pi/3,2\pi/3]~\mathrm{rad}$ & B1, B2 \\
\hline
Open-loop jink & Number of jinks$^{*}$ & $\operatorname{UnifInt}\{1,\ldots,3\}$ & B2$^{*}$ \\
& Jink duration$^{*}$ & $[3,8]~\mathrm{s}$ & B2$^{*}$ \\
& Jink turn-rate fraction$^{*}$ & $[0.5,0.9]$ of bandit maximum turn rate & B2$^{*}$ \\
\hline
\end{tabular}
\end{table}

\subsection{Performance Metrics \& Bayesian Analysis}
\label{sec:bayesian_metrics}

The effectiveness of guard tactics is evaluated based upon performance metrics described in Table~\ref{tab:performance_metrics}. These attempt to measure how effectively the guards have defended the high-value assets (HVAs or ``ladies'' in Rusnak's terminology~\cite{rusnak2005lady}) and how much control effort was required on the part of the guards. %

\begin{table}[t]
\centering
\caption{Per-trial performance metrics and Bayesian analyses. The reported
posterior intervals are central $95\%$ credible intervals.}
\label{tab:performance_metrics}
\small
\renewcommand{\arraystretch}{1.2}
\begin{tabular}{p{1.05in} p{1.65in} p{1.40in} p{1.70in}}
\hline
\textbf{Metric} & \textbf{Per-trial observation} & \textbf{Bayesian model and prior} & \textbf{Posterior quantities reported} \\
\hline
defense-success
& $Y_{\mathrm{sd}} \in \{0,1\}$, equal to one only if no asset is reached and every bandit is intercepted; zero otherwise
& Bernoulli likelihood with $\theta_{\mathrm{sd}} \sim \mathrm{Beta}(1,1)$
& $\mathbb{E}[\theta_{\mathrm{sd}}\mid D]$ and its $95\%$ credible interval estimated from $100{,}000$ posterior samples; higher is better \\
\hline
HVAs-reached
& $S_{\mathrm{lr}} \in [0,1]$, the fraction of assets reached in one trial
& One hundred equal-width bins on $[0,1]$; multinomial likelihood with symmetric $\mathrm{Dirichlet}(1/K,\ldots,1/K)$ prior for $K=100$ bins (total concentration one)
& $\mathbb{E}[\mu_{\mathrm{lr}}\mid D]$ and its $95\%$ credible interval estimated from $100{,}000$ posterior samples; lower is better \\
\hline
Bandits intercepted
& $S_{\mathrm{bi}} \in [0,1]$, the fraction of bandits intercepted in one trial
& One hundred equal-width bins on $[0,1]$; multinomial likelihood with symmetric $\mathrm{Dirichlet}(1/K,\ldots,1/K)$ prior for $K=100$ bins (total concentration one)
& $\mathbb{E}[\mu_{\mathrm{bi}}\mid D]$ and its $95\%$ credible interval estimated from $100{,}000$ posterior samples; higher is better \\
\hline
mean-control-saturation
& $S_{\mathrm{sat}} \in [0,1]$, the mean across guards of the fraction of recorded active control steps on which the requested action is clipped
& One hundred equal-width bins on $[0,1]$; multinomial likelihood with symmetric $\mathrm{Dirichlet}(1/K,\ldots,1/K)$ prior for $K=100$ bins (total concentration one)
& $\mathbb{E}[\mu_{\mathrm{sat}}\mid D]$ and its $95\%$ credible interval estimated from $100{,}000$ posterior samples; lower is better \\
\hline
\end{tabular}
\end{table}

The boolean \emph{defense-success} outcome is of primary concern for our performance analysis. For a given Monte Carlo trial episode, defense-success $Y_{\mathrm{sd}}$ is True if and only if all bandits are intercepted and no HVAs are reached. 
For statistical analysis, a Beta-Binomial model \cite[Ch.~2]{gelman2013bayesian} is applied whereby defense-success is treated as a Bernoulli trial and we seek to estimate the probability of successful defense for a particular set of guard tactics. We use a Beta distribution to model the success probability parameter, $\theta_{\text{sd}}$, and a noninformative (uniform) prior on the distribution.

The \emph{HVAs-reached} ($S_{\mathrm{lr}} \in [0,1]$) and \emph{bandits-intercepted} ($S_{\mathrm{bi}} \in [0,1]$) trial outcomes are real-valued scalars that measure, respectively, the percent of assets reached by bandits and percent of bandits intercepted by guards within a simulated episode. %
For statistical analysis, a Dirichlet-Multinomial model \cite[Ch.~3]{gelman2013bayesian} is applied to these outcomes whereby the real-valued outcomes, $S_{\mathrm{lr}}$ and $S_{\mathrm{bi}}$, are binned into $K$-bins evenly distributed in range $[0,1]$. %
Then Monte Carlo trial data is used to estimate the posterior distribution of the probability mass across the bins. %
A weak, symmetric Dirichlet prior is used.

Note that, while defense-success and bandits-intercepted outcomes are related, there are salient differences between the two. Beyond the fact that one is a boolean while the other is a real-valued scalar, you can get a very different perception of guard tactics effectiveness from these two values. %
For example, if there were a set of guard tactics that \emph{always} intercepted 99 out of 100 bandits within an engagement, then you would have bandit-intercepted rate of $99\%$ but a defense-success rate of $0\%$ because in no trial do you ever catch all of the bandits. 

The \emph{mean-control-saturation} ($S_{\mathrm{sat}} \in [0,1]$) trial outcome measures the proportion of time steps in which any guard agent's control is saturated (i.e. the control command exceeds the actuator limits of the guard vehicle and must be clipped to the bounds), averaged across all guard agents using the same tactics. 
A Dirichlet-Multinomial with a weak, symmetric prior is similarly applied to estimate the posterior distribution of mean-control-saturation.

As previously mentioned, the differential game (DG) and coverage-aware (CA) experiment campaigns use identical ordered trial realizations, i.e. the same random number seed. %
We can use this information to more directly answer questions such as: \emph{what is the probability that the differential game tactics have a higher defense-success probability than coverage-aware tactics?} %
To do so, we define the four possible defense-success outcomes of a paired trial: both DG and CA are successful (1,1), only DG succeeds (1,0), only CA succeeds (0,1), and both fail (0,0). %
By counting the paired outcomes across DG and CA campaigns, we form the Dirichlet posterior distribution as
\begin{equation}
    \begin{aligned}
        (\pi_{11},\pi_{10},\pi_{01},\pi_{00}) \mid D
        &\sim \operatorname{Dirichlet}\!\left(
            n_{11}+\tfrac{1}{2},\ n_{10}+\tfrac{1}{2},\
            n_{01}+\tfrac{1}{2},\ n_{00}+\tfrac{1}{2}
        \right)
    \end{aligned}
    \label{eqn:paired_defense_success_posterior}
\end{equation}
where $n_{ab}$ is the paired-trial count and $\pi_{ab}$ is the joint probability of DG outcome $a$ and CA outcome $b$. %
By recognizing that $\theta_{\mathrm{sd}}^{\mathrm{DG}} = \pi_{11}+\pi_{10}$ and $\theta_{\mathrm{sd}}^{\mathrm{CA}} = \pi_{11}+\pi_{01}$, we can sample the distribution in Eq.~\ref{eqn:paired_defense_success_posterior} to estimate $\Pr(\theta^{\mathrm{DG}}_{\mathrm{sd}} > \theta^{\mathrm{CA}}_{\mathrm{sd}} \mid D^{\mathrm{DG}}, D^{\mathrm{CA}})$.

\section{Results \& Discussion}
\label{sec:results}

Tables~\ref{tab:results_b1_ic1} and \ref{tab:results_b2_ic1} report the Bayesian posteriors of performance metrics (Table~\ref{tab:performance_metrics}) for the non-evasive (B1) and evasive (B2) intruder environments, respectively. Generally speaking, we seek high expected values for defense-success and bandits-intercepted ($\theta_{\mathrm{sd}}$ and $\mu_{\mathrm{bi}}$) and low expected values for HVAs-reached and control saturation ($\mu_{\mathrm{lr}}$ and $\mu_{\mathrm{sat}}$).

\begin{table}[htbp]
    \centering
    \caption{Bayesian performance summary for \textbf{non-evasive bandit tactics}.
    Each entry gives the posterior mean and central $95\%$ credible interval
    (in brackets). The nearest-bandit baseline uses 1,000 Monte Carlo trials;
    the coverage-aware baseline and differential-game tactics each use 500.
    Posterior summaries use 100,000 samples.}
    \label{tab:results_b1_ic1}
    \resizebox{\textwidth}{!}{%
    \renewcommand{\arraystretch}{3.0}
    \begin{tabular}{lcccc}
        \hline
        Guard tactic &
        \shortstack{Defense success\\$\mathbb{E}[\theta_{\mathrm{sd}} \mid D]$} &
        \shortstack{HVAs reached\\$\mathbb{E}[\mu_{\mathrm{lr}} \mid D]$} &
        \shortstack{Bandits intercepted\\$\mathbb{E}[\mu_{\mathrm{bi}} \mid D]$} &
        \shortstack{Control saturation\\$\mathbb{E}[\mu_{\mathrm{sat}} \mid D]$} \\
        \hline
        Nearest-bandit (NB) &
        \shortstack{0.157\\$[0.135, 0.180]$} &
        \shortstack{0.182\\$[0.172, 0.194]$} &
        \shortstack{0.828\\$[0.821, 0.835]$} &
        \shortstack{0.007\\$[0.007, 0.009]$} \\
        Coverage-aware (CA) &
        \shortstack{0.958\\$[0.939, 0.974]$} &
        \shortstack{0.013\\$[0.010, 0.018]$} &
        \shortstack{0.990\\$[0.987, 0.992]$} &
        \shortstack{0.008\\$[0.007, 0.011]$} \\
        Differential-game (DG) &
        \shortstack{\textbf{0.966}\\$\mathbf{[0.949, 0.980]}$} &
        \shortstack{0.011\\$[0.008, 0.016]$} &
        \shortstack{0.991\\$[0.988, 0.993]$} &
        \shortstack{0.006\\$[0.005, 0.009]$} \\
        \hline
    \end{tabular}
    }
    \par\smallskip
    {\textit{Paired DG--coverage-aware comparison (500 trials):}
    $\Pr(\theta^{\mathrm{DG}}_{\mathrm{sd}} > \theta^{\mathrm{CA}}_{\mathrm{sd}} \mid
    D^{\mathrm{DG}}, D^{\mathrm{CA}})=0.858$
    }
\end{table}

\begin{table}[htbp]
    \centering
    \caption{Bayesian performance summary for \textbf{evasive bandit tactics}.
    Each entry gives the posterior mean and central $95\%$ credible interval
    (in brackets). The nearest-bandit baseline uses 1,000 Monte Carlo trials;
    the coverage-aware baseline and differential-game tactics each use 500.
    Posterior summaries use 100,000 samples.}
    \label{tab:results_b2_ic1}
    \resizebox{\textwidth}{!}{%
    \renewcommand{\arraystretch}{3.0}
    \begin{tabular}{lcccc}
        \hline
        Guard tactic &
        \shortstack{Defense success\\$\mathbb{E}[\theta_{\mathrm{sd}} \mid D]$} &
        \shortstack{HVAs reached\\$\mathbb{E}[\mu_{\mathrm{lr}} \mid D]$} &
        \shortstack{Bandits intercepted\\$\mathbb{E}[\mu_{\mathrm{bi}} \mid D]$} &
        \shortstack{Control saturation\\$\mathbb{E}[\mu_{\mathrm{sat}} \mid D]$} \\
        \hline
        Nearest-bandit (NB) &
        \shortstack{0.146\\$[0.125, 0.168]$} &
        \shortstack{0.192\\$[0.181, 0.203]$} &
        \shortstack{0.802\\$[0.794, 0.810]$} &
        \shortstack{0.009\\$[0.008, 0.010]$} \\
        Coverage-aware (CA) &
        \shortstack{0.946\\$[0.925, 0.964]$} &
        \shortstack{0.015\\$[0.011, 0.019]$} &
        \shortstack{0.990\\$[0.986, 0.992]$} &
        \shortstack{0.009\\$[0.008, 0.013]$} \\
        Differential-game (DG) &
        \shortstack{\textbf{0.968}\\$\mathbf{[0.951, 0.982]}$} &
        \shortstack{0.011\\$[0.009, 0.016]$} &
        \shortstack{0.991\\$[0.988, 0.993]$} &
        \shortstack{0.006\\$[0.005, 0.009]$} \\
        \hline
    \end{tabular}
    }
    \par\smallskip
    {\textit{Paired DG--coverage-aware comparison (500 trials):}
    $\Pr(\theta^{\mathrm{DG}}_{\mathrm{sd}} > \theta^{\mathrm{CA}}_{\mathrm{sd}} \mid
    D^{\mathrm{DG}}, D^{\mathrm{CA}})=0.999$
    }
\end{table}

The key result from Tables \ref{tab:results_b1_ic1} and  \ref{tab:results_b2_ic1} is that the differential game tactics produce the highest probability of successful defense of high-value assets ($\mathbb{E}[\theta_{\mathrm{sd}} \mid D]$); higher than both the naive nearest-bandit (NB) and more sophisticated coverage-aware (CA) baseline tactics that ignore the game-theoretic modeling of the problem, instead treating it as a unilateral optimization (Eq.~\eqref{eqn:swarm_defense_optim_problem}). %
The importance of the game modeling is further emphasized when comparing results between Tables \ref{tab:results_b1_ic1}, where the bandits were non-evasive, and  \ref{tab:results_b2_ic1}, where bandits performed evasive maneuvers: we see a modest drop in successful defense when bandits employ evasive maneuvers for the non-game-theoretic baseline tactics (NB and CA); whereas the differential game (DG) tactics are able to maintain their performance. 

Tables \ref{tab:results_b1_ic1} and \ref{tab:results_b2_ic1} also help illustrate the importance of intercepting \emph{all} bandits within an engagement, versus simply intercepting a large majority per engagement. %
In the case of evasive bandits, we see that the nearest-bandit (NB) guard tactics are capable of intercepting an average of 80.2\% of bandits; however, this only results in a 14.6\% percent chance that \emph{all} bandits are intercepted during a particular engagement. %
In the real world, it would be unacceptable to have such a low chance of defending all high-value assets. %
We see how the coverage-aware (CA) guard tactics significantly improve upon this key metric, raising the probability of successfully defending \emph{all} HVAs to 94.6\%. 
However, the differential game (DG) tactics further improve on this key metric, raising the defense success probability to 96.8\% in the presence of evasive bandits. %
Another way to frame this is that the non-game-theoretic CA tactics have a 5.4\% \emph{gap to perfect defense}, and by employing game-theoretic tactics, 40.7\% of this gap may be closed.

Given the relatively high success rate of both the DG and CA guard tactics, it may be asked if differential game tactics are probabilistically ``better'' than the non-game-theoretic baselines, or if these differences may be statistically insignificant. %
To answer this question, Tables~\ref{tab:results_b1_ic1} and \ref{tab:results_b2_ic1} also provide results for the paired-trial posterior analysis described in Sec.~\ref{sec:bayesian_metrics} and show that the probability that DG tactics have a higher probability of successful defense than CA tactics is 85.8\% for non-evasive bandits, rising to 99.9\% for evasive bandits.

\begin{figure}[p]
    \centering
    \begin{minipage}[t]{0.48\textwidth}
        \centering
        \includegraphics[width=\linewidth]{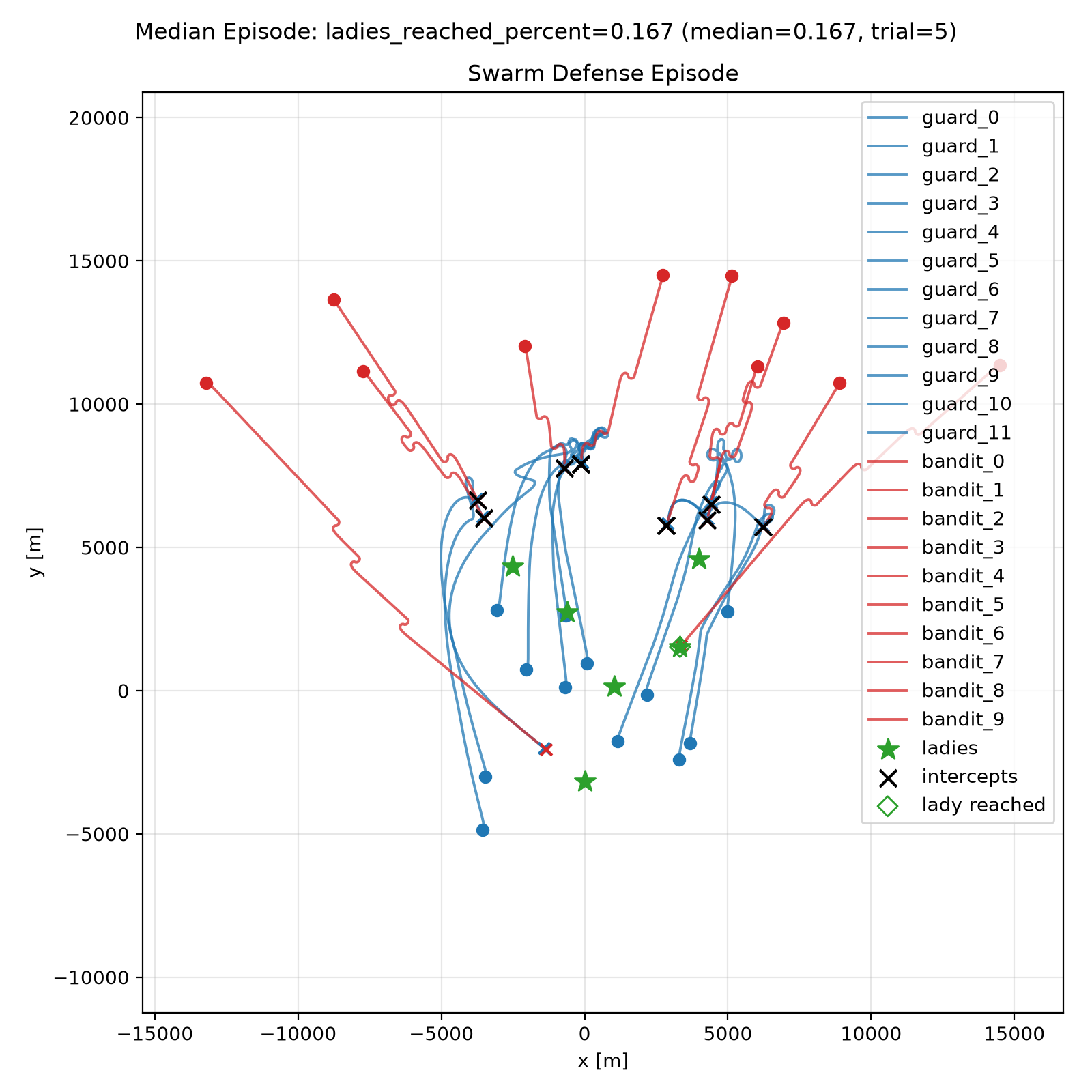}
        \par\smallskip
        \textbf{(a)} Nearest-bandit baseline
    \end{minipage}\hfill
    \begin{minipage}[t]{0.48\textwidth}
        \centering
        \includegraphics[width=\linewidth]{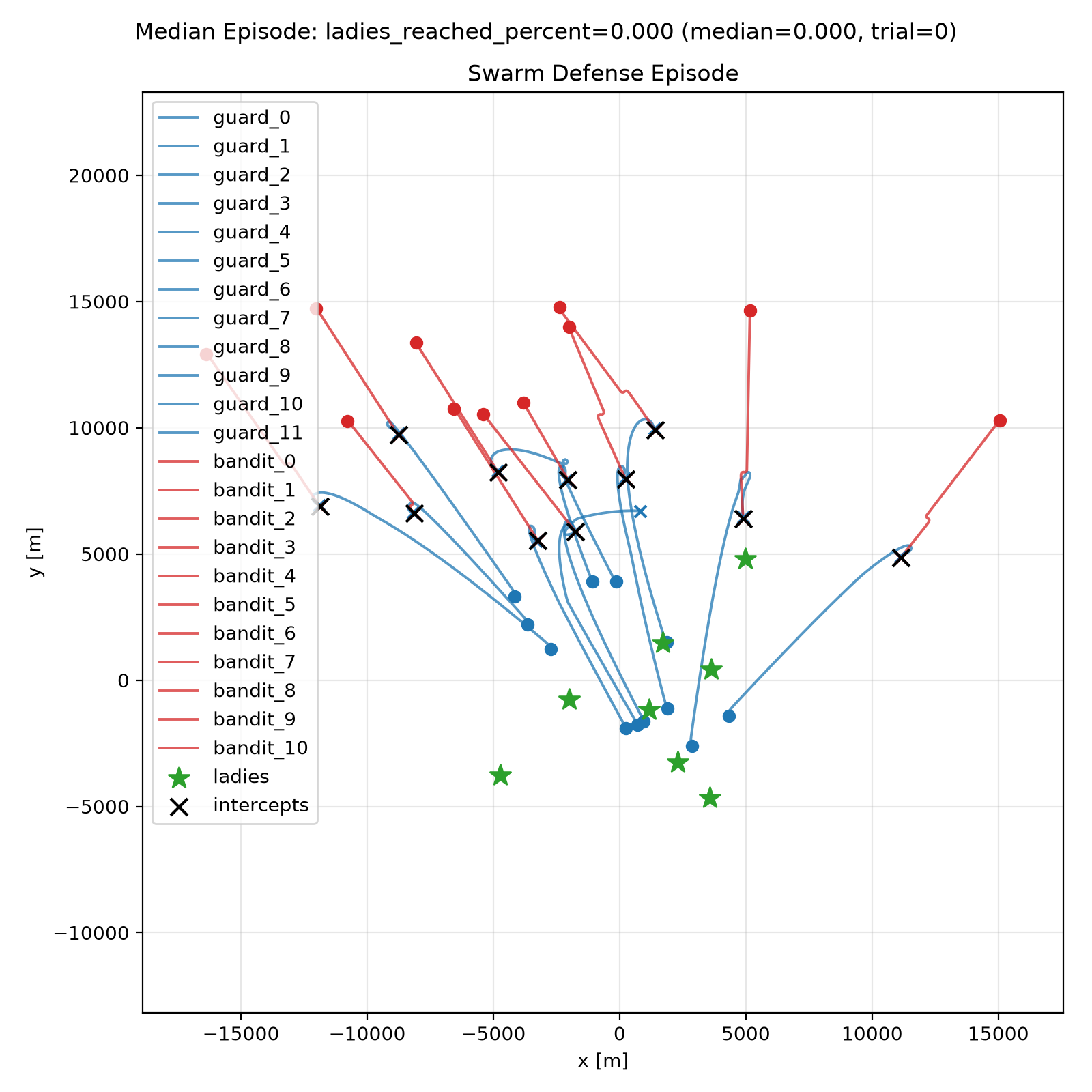}
        \par\smallskip
        \textbf{(b)} Coverage-aware baseline
    \end{minipage}

    \par\medskip
    \begin{minipage}[t]{0.48\textwidth}
        \centering
        \includegraphics[width=\linewidth]{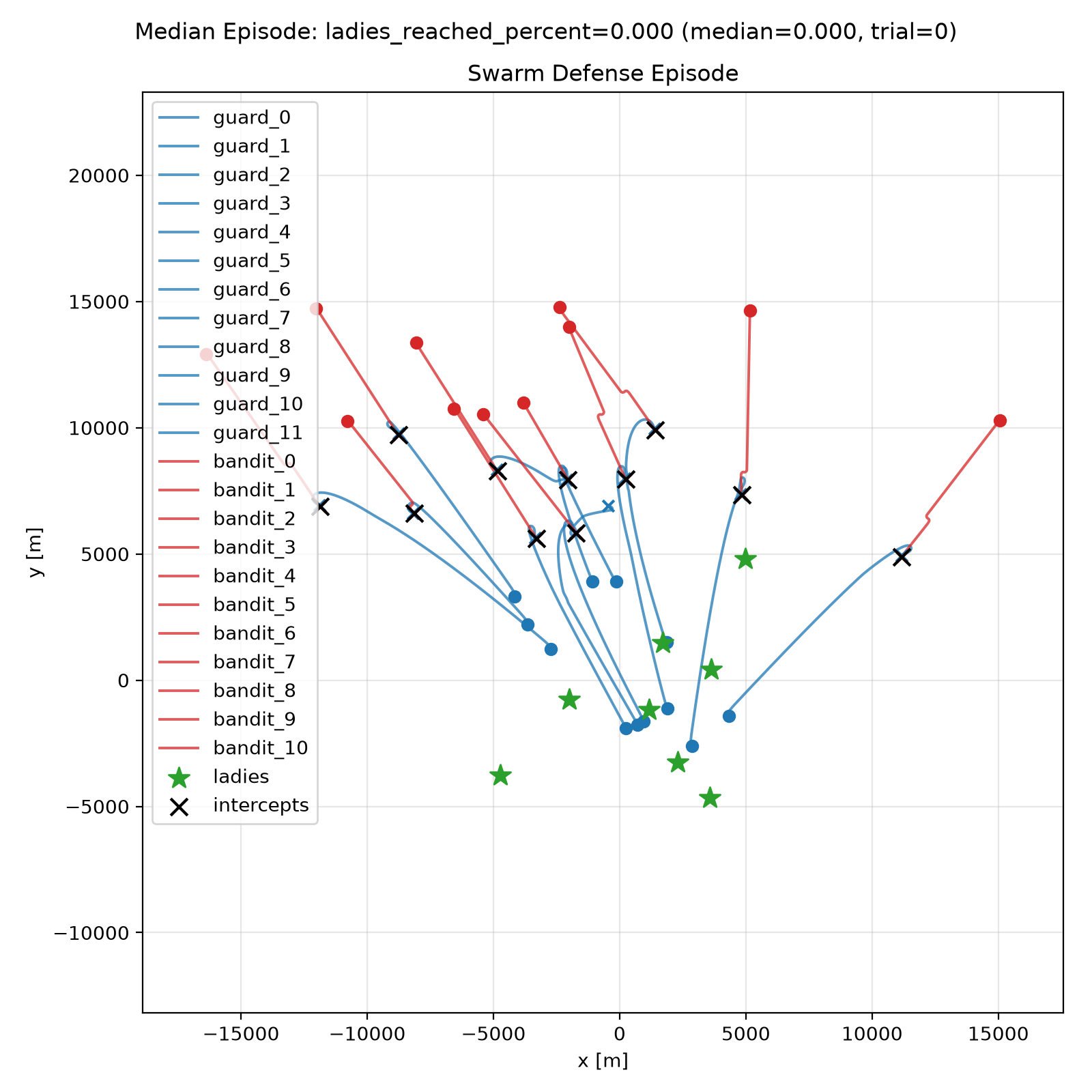}
        \par\smallskip
        \textbf{(c)} Differential-game tactics
    \end{minipage}
    \caption{Median evasive-bandit (B2) episode selected independently for each tactic by
    HVA-reached percentage: (a) nearest-bandit baseline (0.167), (b)
    coverage-aware baseline (0.000), and (c) differential-game tactics
    (0.000).}
    \label{fig:b2_median_ladies_reached}
\end{figure}

\begin{figure}[p]
    \centering
    \begin{minipage}[t]{0.48\textwidth}
        \centering
        \includegraphics[width=\linewidth]{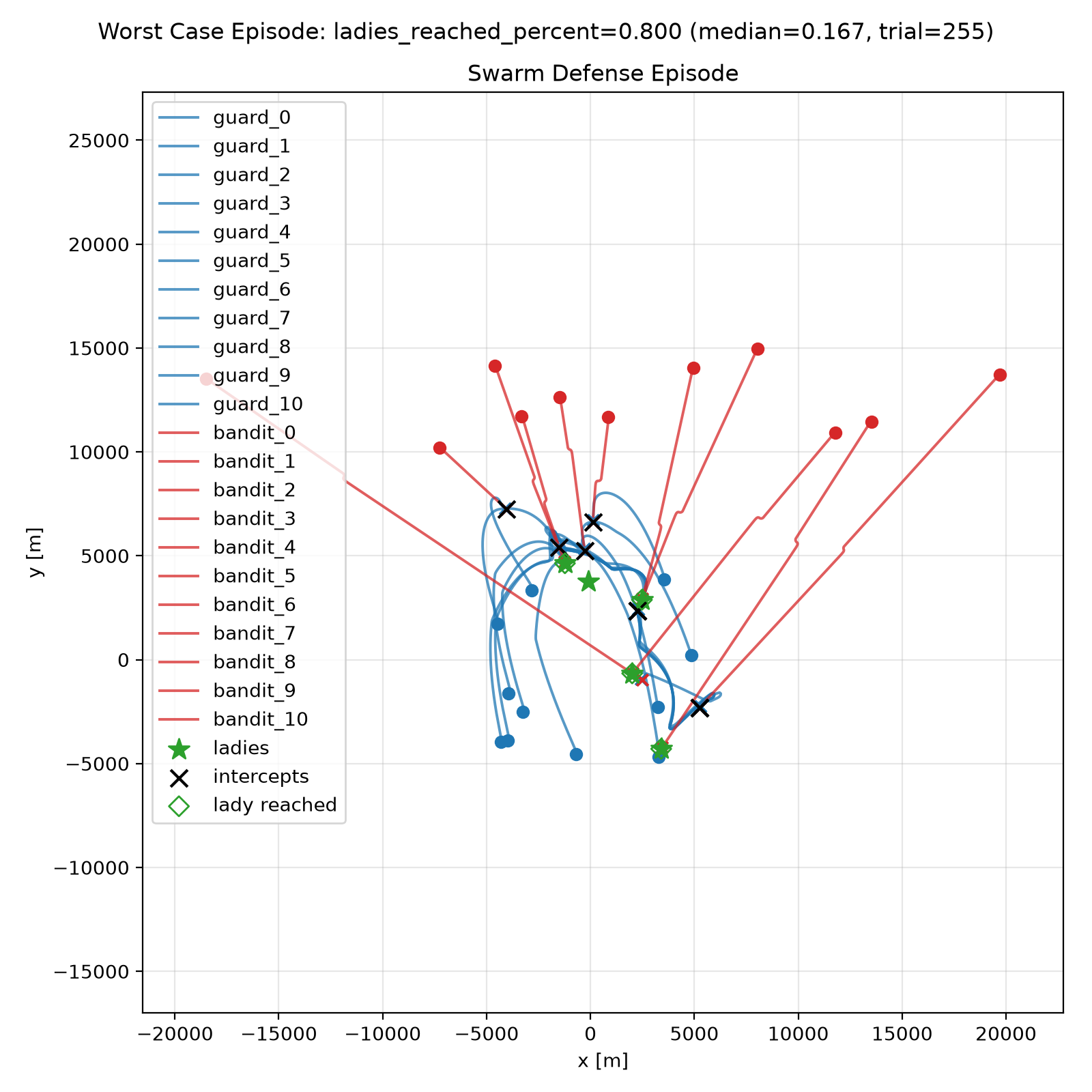}
        \par\smallskip
        \textbf{(a)} Nearest-bandit baseline
    \end{minipage}\hfill
    \begin{minipage}[t]{0.48\textwidth}
        \centering
        \includegraphics[width=\linewidth]{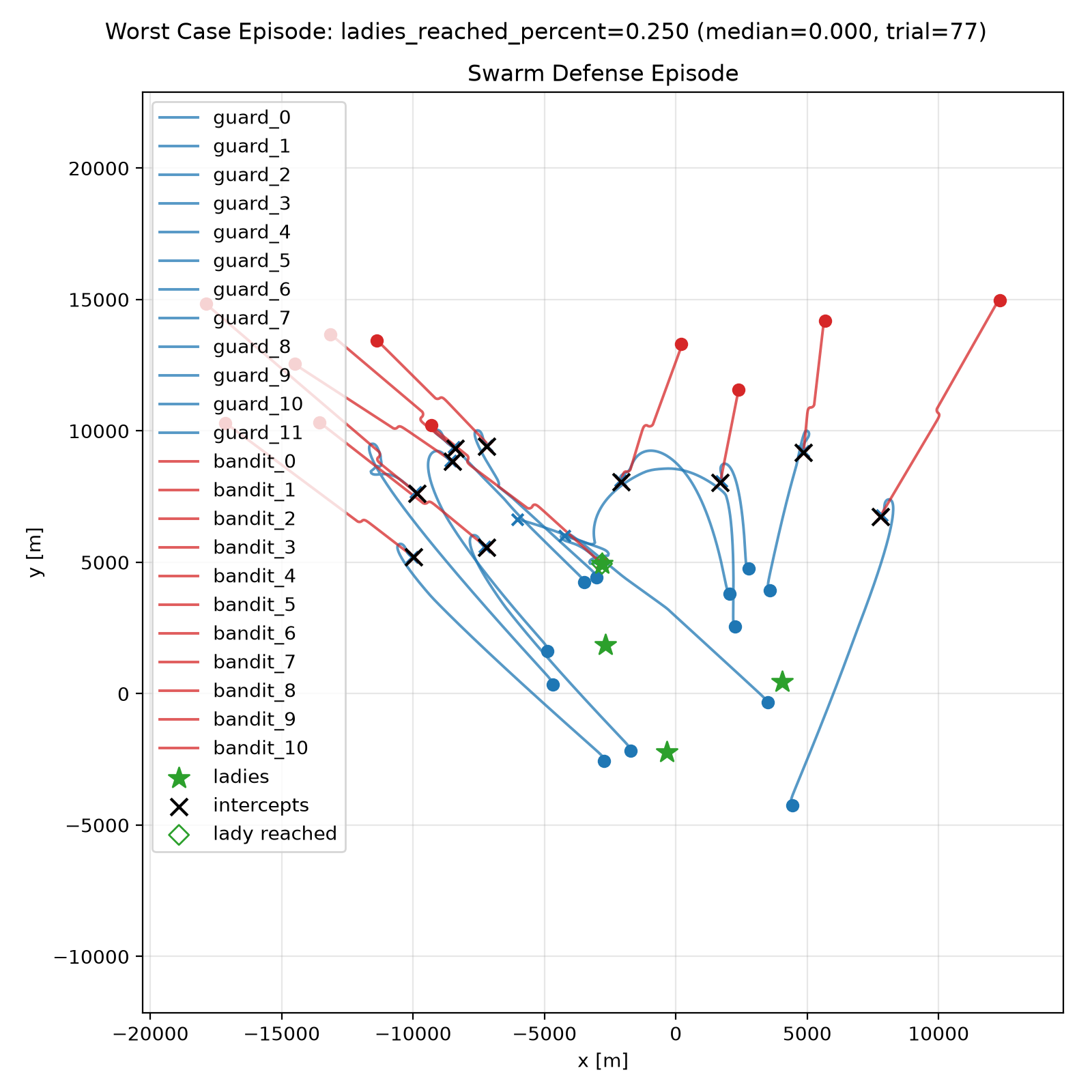}
        \par\smallskip
        \textbf{(b)} Coverage-aware baseline
    \end{minipage}

    \par\medskip
    \begin{minipage}[t]{0.48\textwidth}
        \centering
        \includegraphics[width=\linewidth]{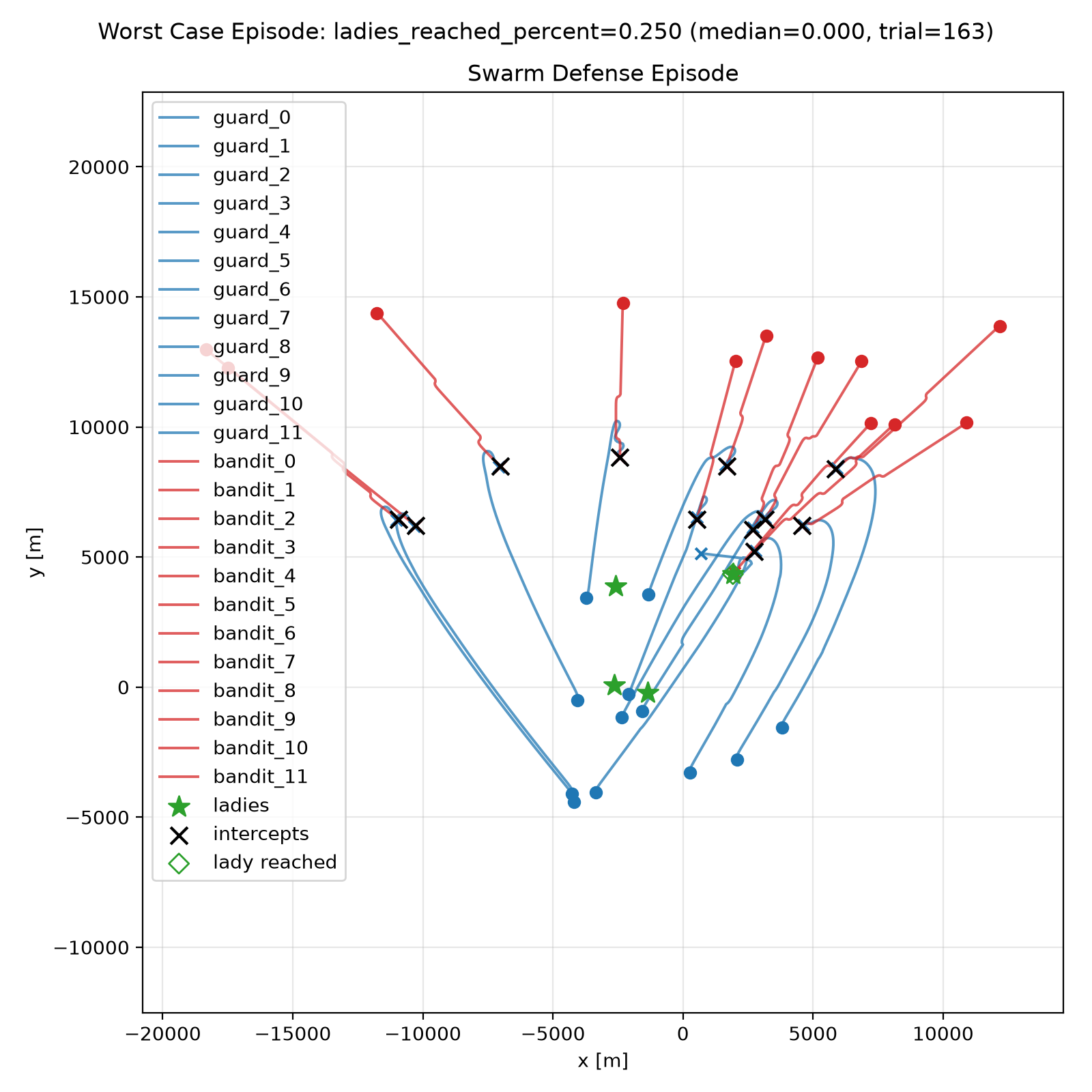}
        \par\smallskip
        \textbf{(c)} Differential-game tactics
    \end{minipage}
    \caption{Worst observed evasive-bandit (B2) episode selected independently for each
    tactic by HVA-reached percentage: (a) nearest-bandit baseline (0.800),
    (b) coverage-aware baseline (0.250), and (c) differential-game tactics
    (0.250).}
    \label{fig:b2_worst_ladies_reached}
\end{figure}

Figures~\ref{fig:b2_median_ladies_reached} and \ref{fig:b2_worst_ladies_reached} illustrate the vehicle trajectories for a selection of evasive-bandit (B2) episodes with the nearest-bandit (NB), coverage-aware (CA), and differential-game (DG) guard tactics. %
Figure~\ref{fig:b2_median_ladies_reached} represents a median trial in each of the three campaigns as measured by the percentage of HVAs-reached by bandits; Figure~\ref{fig:b2_worst_ladies_reached} represents the worst trial from each campaign. %
These figures are presented, mostly, to give intuition of what an episode rollout would look like and how guards distribute themselves across bandits. 
It is difficult to discern meaningful differences in the behavior of the CA and DG tactics within these individual trials, which is why the statistical results in Tables~\ref{tab:results_b1_ic1} and \ref{tab:results_b2_ic1} are the more salient for our analysis.

\clearpage

\section{Limitations \& Future Work}
\label{sec:out_of_scope}

There are a number of limitations to this current work that need addressing in future work in order to mature game-theoretic algorithms for use in real-world swarm defense. %
It has been assumed that all guard drones---or at least a centralized guard coordinator---may observe the state of all drones and assets at all times. %
It has been assumed that a centralized guard coordinator may communicate with all guard drones at all times, with no bandwidth limitations. %
A highly simplified aircraft model has been employed; neglecting even variations in altitude, let alone high-fidelity aerodynamics and flight control. %
The interception adjudication model is also highly simplified. Sensor models, such as cameras and accelerometers, would be needed to improve simulation of interception dynamics. 

While we have modeled both non-evasive and evasive (i.e. pre-scripted, open-loop) bandits, we have not yet simulated \emph{truly reactive} bandits which are able to perceive the state of guard drones and maneuver accordingly to better evade interception.

We have also limited our experiments to those in which guard drones are faster than bandits; situations with faster bandits have been left out of the scope of this paper for several reasons. %
First, bandits do not perceive or react to guards; therefore, guards are incapable of any kind of ``shepherding'' of bandits, regardless of relative speeds. %
This means that all interceptions of faster-bandits must be done head-on or cross-range so as to achieve a positive closure speed; however, these geometries will lead to very high relative velocities (e.g. 100s of meters per second). %
For extremely high relative velocities, the current modeling of interception adjudication is unrealistic and any simulation results produced would just be artifacts of these unrealistic simulation assumptions; not reflections of real-world performance. %
Extensive, high-fidelity platform and sensor models would be needed to make them realistic. In other words, the only thing an overmatched guard team could do is throw many interceptors head-first at a fast bandit until one of them happens to achieve an ``interception'', but that simulated interception would not be based on any realistic physics; and thus there is nothing to be learned from this exercise.

\section*{Acknowledgements}
\label{sec:acknowledgements}

The authors thank David Fridovich-Keil and Kushagra Gupta at The University of
Texas at Austin for helpful discussions and feedback on this work.

\clearpage
\appendix
\section{Appendices}

\subsection{Discounted-Redundancy Assignment Reformulation}
\label{app:discounted_assignment}

The discounted-redundancy objective in Sec.~\ref{sec:defend_target_assignment}
can be represented as a rectangular linear assignment problem. For every active
bandit $j$, create $N_{G_A}$ virtual assignment slots $(j,k)$, with
$k\in\{1,\ldots,N_{G_A}\}$. The $k$th slot represents the $k$th guard assigned to
bandit $j$ and has marginal value $v_j\beta^{k-1}$. Let
$y_{i,j,k}\in\{0,1\}$ indicate whether guard $i$ is assigned to slot $(j,k)$,
and define the corresponding linear cost as
\begin{equation}
    d^{\mathrm{TA}}_{i,j,k}
    = c^{\mathrm{TA}}_{ij} - v_j\beta^{k-1}.
\end{equation}
The resulting assignment problem is
\begin{align}
    \min_{y_{i,j,k}} \quad &
    \sum_{i\in G_A}
    \sum_{j\in B_A}
    \sum_{k=1}^{N_{G_A}}d^{\mathrm{TA}}_{i,j,k}y_{i,j,k},
    \\
    \text{subject to} \quad &
    \sum_{j\in B_A}\sum_{k=1}^{N_{G_A}}y_{i,j,k}=1,
    \qquad \forall i\in G_A,
    \\
    & \sum_{i\in G_A}y_{i,j,k}\leq1,
    \qquad \forall j\in B_A,\;
    k\in\{1,\ldots,N_{G_A}\},
    \\
    & y_{i,j,k}\in\{0,1\}.
\end{align}
There are $N_{G_A}$ guard rows and $N_{B_A}N_{G_A}$ virtual bandit-slot columns. Although
stated with binary variables, the linear-program relaxation is integral because
the bipartite assignment constraint matrix is totally unimodular. Since slot
values weakly decrease with $k$, an optimal solution can always be interpreted
as filling earlier slots for a bandit before later slots. The cost matrix can be
solved directly using SciPy's
\href{https://docs.scipy.org/doc/scipy/reference/generated/scipy.optimize.linear_sum_assignment.html}{\texttt{scipy.optimize.linear\_sum\_assignment}}
routine.

\nocite{fridovichkeil2020efficient,lecleach2022algames}
\bibliographystyle{plain}
\bibliography{references}

\end{document}